\documentclass[submission]{jsfds} 
\usepackage[latin1]{inputenc} 
\usepackage[T1]{fontenc}
\usepackage{amsmath}
\usepackage{amssymb}
\usepackage{graphicx}
\usepackage{cite}
\usepackage{color} 
\usepackage[labelfont=bf,labelsep=period,justification=raggedright]{caption}
\usepackage{url}

\setmainlanguageenglish
\startlocaldefs % for example

 \def\1{1\kern-.20em {\rm l}} 
\endlocaldefs

\date{}

\begin{document}
\begin{frontmatter}
%\verb+\arxiv{math.PR/00000000}+ %if available, 
\title{Defining a robust biological prior from Pathway Analysis to drive Network Inference.}
\runtitle{Network Inference from a biological prior} 
\alttitle{Construction d'un a priori biologique robuste à partir de l'analyse de voies m\'etaboliques pour l'inférence de réseaux.}
\begin{aug} 
\auteur{%
 \prenom{Marine} 
\nom{Jeanmougin}
\thanksref{t1,t2}
\contact[label=e1]{marine.jeanmougin@genopole.cnrs.fr}} \and%
\auteur{% 
\prenom{Mickael}
\nom{Guedj} \thanksref{t2} 
\contact[label=e2]{mickael.guedj@pharnext.com}} \and% 
\auteur{%
\prenom{Christophe}
 \nom{Ambroise} \thanksref{t1}
 \contact[label=e3]{Christophe.Ambroise@genopole.cnrs.fr}}
\affiliation[t1]{Statistics and Genome laboratory UMR CNRS 8071, University of Evry, Evry, France.\\ 
\printcontact{e1}}
\affiliation[t2]{Department of Biostatistics, Pharnext, Paris, France.\\ 
\printcontact{e2} and \printcontact{e3}}
\runauthor{Jeanmougin, Guedj and Ambroise} 
\end{aug}

\begin{abstract}\\
Inferring genetic networks from gene expression data is one of the most challenging work in the post-genomic era, partly due to the vast space of possible networks and the relatively small amount of data available. In this field, Gaussian Graphical Model (GGM) provides a convenient framework for the discovery of biological networks.\\
In this paper, we propose an original approach for inferring gene regulation networks using a robust biological prior on their structure in order to limit the set of candidate networks.\\

Pathways, that represent biological knowledge on the regulatory networks, will be used as an informative prior knowledge to drive Network Inference. This approach is based on the selection of a relevant set of genes, called the ``molecular signature'', associated with a condition of interest (for instance, the genes involved in disease development). In this context, differential expression analysis is a well established strategy. However outcome signatures are often not consistent and show little overlap between studies. Thus, we will dedicate the first part of our work to the improvement of the standard process of biomarker identification to guarantee the robustness and reproducibility of the molecular signature.\\

Our approach enables to compare the networks inferred between two conditions of interest (for instance case and control networks) and help along the biological interpretation of results. Thus it allows to identify differential regulations that occur in these conditions.
We illustrate the proposed approach by applying our method to a study of breast cancer's response to treatment.
 \end{abstract}

 \begin{altabstract}\\ 
% and its translation in french 
L'inférence de réseaux génétiques à partir de données issues de biopuces est un des défis majeurs de l'ère post-génomique, en partie à cause du grand nombre de réseaux possibles et de la quantité relativement faible de données disponibles. Dans ce contexte, la théorie des modèles graphiques gaussiens est un outil efficace pour la reconstruction de réseaux.\\
A travers ce travail nous proposons une approche d'inférence de réseaux de régulation à partir d'un \textit{a priori} biologique robuste sur la structure des réseaux afin de limiter le nombre de candidats possibles.\\

Les voies métaboliques, qui rendent compte des connaissances biologiques des réseaux de régulation, nous permettent de définir cet \textit{a priori}. Cette approche est basée sur la sélection d'un ensemble de gènes pertinents, appelé ``signature moléculaire'', potentiellement associé à un phénotype d'intérêt (par exemple les gènes impliqués dans le développement d'une pathologie). Dans ce contexte, l'analyse différentielle est la strategie prédominante. Néanmoins les signatures de gènes diffèrent d'une étude à l'autre et la robustesse de telles approches peut être remise en question. Ainsi, la première partie de notre travail consistera en l'amélioration de la stratégie d'identification des gènes les plus informatifs afin de garantir la robustesse et la reproductibilité de la signature moléculaire.\\

Notre approche vise à comparer les réseaux inférés dans différentes conditions d'étude et à faciliter l'interprétation biologique des résultats. Ainsi, elle permet de mettre en avant des régulations différentielles entre ces conditions.\\
Nous appliquerons notre méthode à l'étude de la réponse au traitement dans le cancer du sein.
\end{altabstract}

\begin{altkeywords}
\mot{Inférence de réseaux}
\mot{Modèle graphique gaussien}
\mot{Pénalisation $\ell_1$}
\mot{Information a priori}%
\mot{Analyse de voies métaboliques}
\end{altkeywords}

\begin{AMSclass}
\kwd{60K35}
\kwd{}
\end{AMSclass}

\end{frontmatter}

\section{Introduction}

Advances in Molecular Biology and substantial improvements in microarray technology have led biologists toward high-throughput genomic studies. It has become possible to detect tens of thousands of genes and compare their expression levels between samples in a single experiment. These data describe dynamic changes in gene expression closely related to regulatory events without offering any explanation on how this is all managed by the genome. The use of microarrays to discover differentially expressed genes between two or more conditions (patients \textit{versus} controls for instance) has found many applications.
These include the identification of disease biomarkers (i.e. the genes potentially associated with a given disease) that may be important in the diagnosis of the different types or subtypes of diseases. A common challenge faced by the researchers is to translate the identified sets of differentially expressed genes into a better understanding of the underlying biological phenomena.\\

A wealth of analysis tools are available among which Pathway Analysis and Network Inference are  getting especially popular in the field of gene expression data analysis. They both hold great promise to interpret genome-wide profile.\\

Pathways can be defined as sets of gene products interacting in order to achieve a specific cellular function (cell cycle or apoptosis for instance). Pathway Analysis aims at determining whether the set of differentially expressed genes is ``enriched'' by a given pathway or cellular function. Pathway Analysis approaches have the advantage of providing a clear and direct biological interpretation. However it can appear restrictive as pathways represent a strong biological prior and results will not deviate from our current knowledge of the cellular mechanisms. \\

On the other side, the purpose of Network Inference is to predict the presence or absence of edges between a set of genes known to form the vertices of a network, the prediction being based on gene expression levels. Various statistical frameworks have been proposed to solve the network inference problem \cite{Jong2002,Werhli2006}. Graphical models such as Bayesian network \cite{Friedman2000} and constraint-based methods are popular frameworks to model the gene regulatory networks. Modelings with the Boolean networks \cite{Kauffman1969,Liang1998,Thomas1973,Remy2008} or differential equations have also been investigated.\\
Discovering structures of regulation networks based on large-scale data represents a major challenge in Systems Biology, partly because the space of possible networks is often too large compared to the relatively small amount of data available. Moreover, exploring all possible network topologies is not computationally feasible.
However, all the networks are not equally plausible and the use of biological knowledge may help to limit the set of candidate networks. Little research have already been made to associate biological informative prior to Network Inference in a Bayesian network framework \cite{Mukherjee2008} or by the use of dynamic Bayesian networks \cite{Bernard2005} from time series data. In the field of Machine Learning other kinds of approaches have been proposed based on kernel metric learning \cite{Yamanishi2004,Vert2005}.\\ 

In this paper, we propose a 3-step approach to define a biological prior on network structure to drive the Network Inference. Our intention here is to overcome both Pathway Analysis and Network Inference limitations by integrating these two approaches. \\
 We will dedicate the first part of our work to the improvement of the standard process of biomarker identification to guarantee the robustness and reproducibility of the gene signature. Indeed due to the biological and technical variability within the expression data, robust gene selection is a crucial issue. Most informative genes have to be identified to ensure the inference of a relevant network.\\
The second step consists in correlating the signature with a priori defined gene sets from biological pathway databases. The resulting metabolic and signaling pathways will be used as an informative prior in the last step.\\
Finally, a network is inferred with the idea to encourage the inference of an edge between two genes when they are known to participate to the same cellular function. A way to describe the interactions among the genome functional elements is by using conditional independencies and, more concretely, graphical models \cite{Pearl1988,Whittaker1990,Lauritzen 1996} which provide a convenient setting for inferring gene regulatory networks. We base our approach on this strategy implemented in the \verb?R? package \verb?SIMoNe? (Statistical Inference for Modular Network) \cite{Chiquet2010} which enables inference of undirected networks based on partial correlation. In this approach, Chiquet et al presents the first method dedicated to the inference of multiple graphs. In other words, when various experimental conditions are available, they propose to estimate multiple GGMs jointly, in a multi-task framework by coupling the estimation of several networks.
\\

We conduct our global approach on breast cancer data to identify the molecular mechanisms underlying the response to treatment using the cohort from Hess et al. 2006 \cite{Hess2006} by comparing the networks inferred under 2 conditions: patients who achieve pathological complete response (pCR) and those who do not (not-pCR).
\section{Materials and Methods}

\begin{figure}[!ht]
\begin{center}
\includegraphics[trim = 50mm 15mm 50mm 30mm ,scale=0.4,clip]{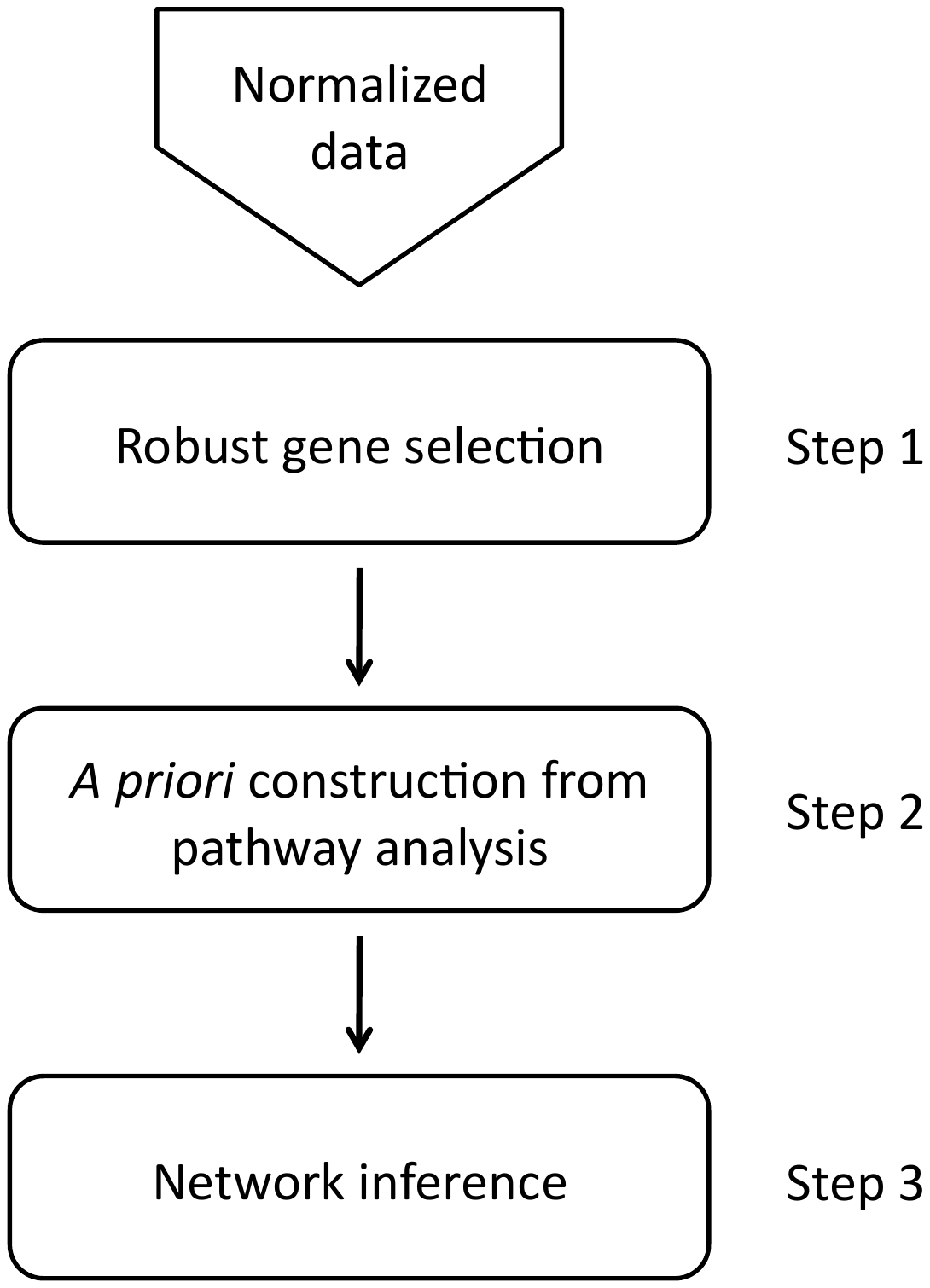}
\end{center}
\caption{\textbf{Global approach of expression data analysis.} Our approach follows a 3-step process: from the expression data we first perform a robust selection to identify the most informative genes, called the signature. Then, we define the biological prior by conducting a Pathway Analysis on the gene signature. The last step consists in inferring a network driven by the prior defined at the previous step.}
\label{fig:PROCESS}
\end{figure}

The purpose of this work is to provide a global framework for transcriptomic data analysis and in particular for inferring co-expression networks from a relevant biological informative prior based on Pathway Analysis. The  co-expression networks are seen  as regulation networks for which the direction of the edges and hence causality events are not known. 
 Such analyses require as a preliminary step to select the genes involved in the phenotype of interest that are assumed to be differentially expressed between the conditions to compare. This selection is of great importance since the results of Pathway Analysis and Network Inference directly depend upon it.\\
Hence, our global approach can be viewed as a 3-step process: i)  an improved selection of differentially expressed genes ii)  a Pathway Analysis and iii)  a Network Inference strategy using Pathway Analysis results as an informative biological prior. These 3 steps are fully described below and summarized in  Figure \ref{fig:PROCESS}. 
 
Let us first introduce some notations. We   consider hereafter  a  sample of a random vectors
$X_{cr}= (X_{cr}^1,\dots,X_{cr}^p)$ where $X_{cr}^i$ is the level of expression observed for gene $i$, replicate $r$, under condition $c$;  The vector $X_{cr}$  is assumed to  be Gaussian with
positive    definite   covariance   matrix    
${\boldsymbol\Sigma^c}   = (\Sigma_{ij}^c )_{(i,j)\in\mathcal{P}^2}$ where  $\mathcal{P}=\{1,\dots,p\}$.
 No  loss  of  generality  is
involved    when   centering    $X_{cr}$,   so    we   may    assume   that
$X_{cr}\sim\mathcal{N}(\mathbf{0}_p,{\boldsymbol\Sigma^c})$.

\subsection*{Step1: Gene selection by robust differential analysis}
Given two conditions to compare (called condition $1$ et condition $2$ in the following), a key question in the first step of expression data analysis is to identify a set of genes that shows a differential expression between the two conditions of interest.
This consists, for each gene, in testing the null hypothesis ($H_0$) that the expected values of expression are equal between the two conditions, against the alternative hypothesis ($H_1$) that they differ. The general model is then given by:
$$\mathbb{E}(X^i_{cr}) = \mu^i_{c}\quad \text{and} \quad Var(X^i_{cr}) = (\sigma^i_{c})^2 .$$
So defined, the null hypothesis to test comes down to: 
\[\left\{ 
\begin{array}{l l l}
 H_0: & \: \mu^i_{1} = \mu^i_{2},\\
 H_1: & \: \mu^i_{1} \neq \mu^i_{2}.\\ \end{array} \right. \]

At first glance,  differential analysis appears  easy enough but in practice results from differential analysis show a lack of reproducibility when compared in the literature \cite{Ein-Dor2005}. In fact, due to the biological and technical noise inherent to high-throughput biological data, methodological aspects are still to investigate to obtain a more robust selection of differentially expressed genes. To do so we propose several methodological improvements to the strategy commonly used, that is generally reduced to the application of a \textit{t}-test. \\

If the $t$-test is certainly the most natural and popular test to assess differential expression, its efficacy in term of variance modeling and power on small sample sizes has been seriously questioned. The recent study from Jeanmougin et al. \cite{Jeanmougin2010} demonstrates that the Bayesian framework proposed by \texttt{limma} \cite{Smyth2004} outperforms all the  usual  methods to test for differential expression. Briefly, \texttt{limma} has the same interpretation as an ordinary $t$-statistic except that the standard errors have been moderated across genes:
$$t^i_{\text{\tiny{limma}}} = \frac{\bar{x}^i_{1\cdot} - \bar{x}^i_{2\cdot}}{S^i\sqrt{\frac{1}{n_1}+\frac{1}{n_2}}},$$

where $n_1$ and $n_2$ are number of replicates of conditions 1 and 2 and $\bar{x}^i_{1\cdot}$ is the average expression level for gene $i$ and condition 1 (across all possible replicates).\\
Indeed, posterior variance, $S^i$, has been substituted into the classical \textit{t}-statistic in place of the usual variance. Using Bayes rules, this posterior variance becomes a combination of an estimate obtained from a prior distribution and the pooled variance. Including a prior distribution on variances has the effect of borrowing information from the ensemble of the genes to aid with inference about each individual gene. Thus, the posterior values shrink the observed variances towards a common value, that is why it is called ``moderated'' \textit{t}-statistic. This approach is implemented in the R package \texttt{limma}.\\

\begin{figure}[!ht]
\begin{center}
\includegraphics[trim = 0 0mm 0 0mm ,scale=0.4,clip]{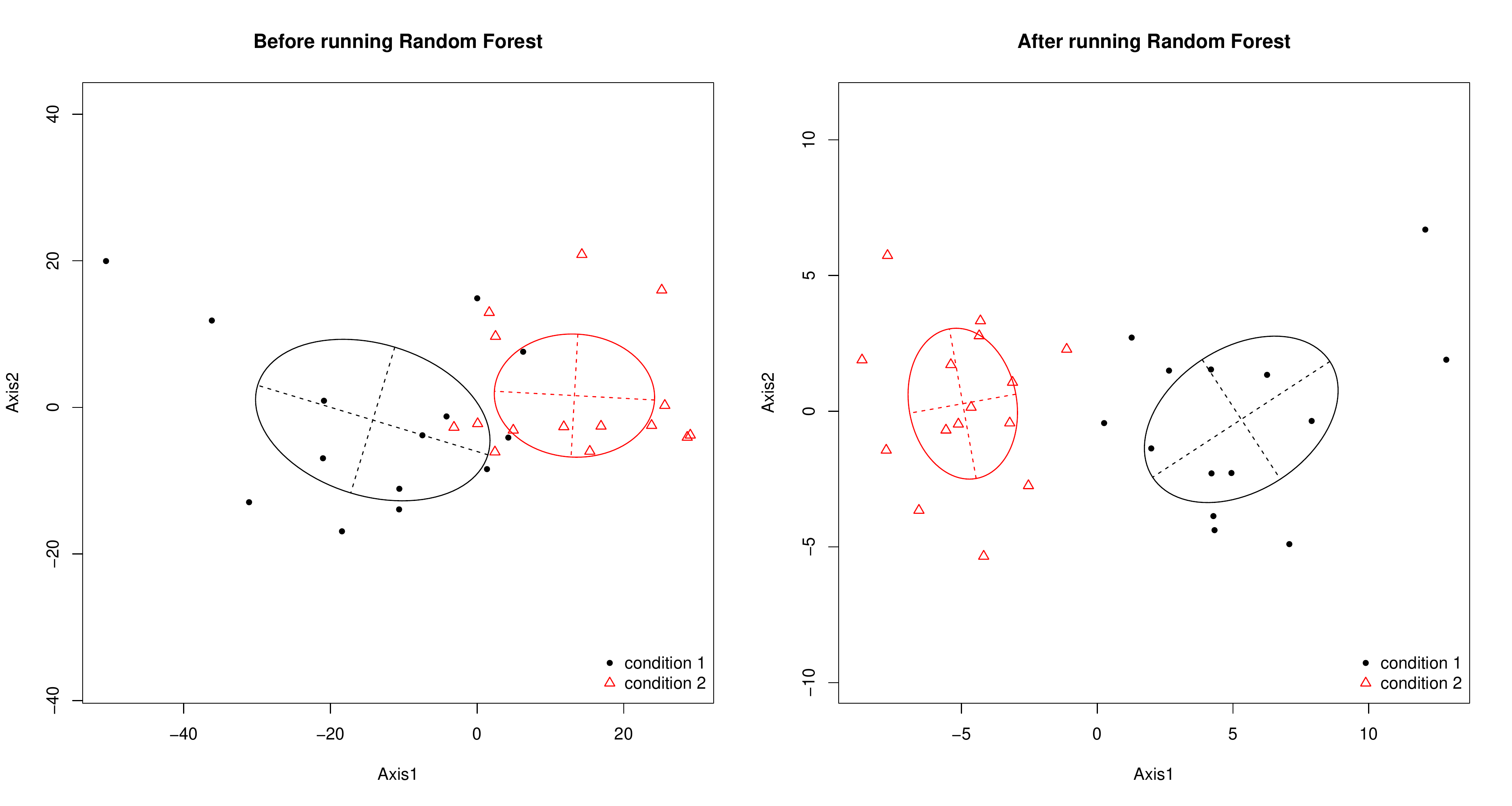}
\end{center}
\caption{\textbf{Improving discrimination between conditions using Random Forests}. A Principal Component Analysis is performed on conditions of interest. The signature obtained by Random Forest (RF) algorithm enables to better discriminate condition 1 from condition 2 as shown in the figures before and after running the RF.}
\label{PCA}
\end{figure}

Despite the use of \texttt{limma}, such a hypothesis testing strategy to select differentially expressed genes between 2 conditions can be biased by outliers. In this context, re-sampling based approaches are pretty common methods. Here, we use random forest \cite{Breiman2001} for improving the overall homogeneity of the signature obtained from differential analysis. The algorithm consists in fitting many binary decision trees built using several bootstrap samples and then in combining the predictions from all the trees. Random forest uses both bagging and random variable selection for tree building. Thus, at each node, a given number of input variables are randomly chosen and the best split is calculated only within this subset. The Out-Of-Bag sample (i.e. the set of observations which are not used for building the current tree) is used to estimate the prediction error and then to evaluate variable importance.\\
We apply Random Forest algorithm (implemented in the R package \verb?randomForest?) to rank the significant genes resulting from \texttt{limma} according to variable importance (i.e ability to well discriminate the conditions). This approach enables to clean the set of differentially expressed genes from putative outliers by selecting most important variables. A simple Principal Component Analysis of the samples before and after the application of the Random Forest algorithm clearly show how  it increases the separation of the conditions (Figure \ref{PCA}).\\

Finally little agreement between studies in the literature is also due to the heterogeneity inherent to genetic diseases and to sampling variation for moderate sample size. Thus, two differential analyses based on two different studies may lead to different selection of genes, however resulting from the same regulatory process. Motivated by the observation that genes causing the same phenotype are likely to be functionally related \cite{Gandhi2006,Lage2007} (i.e they form some kind of module \cite{Oti2007}, for instance a multi-protein complex), we explore a network-based approach to identify modules of genes rather than individual genes. We will refer to this approach as ``functional partners identification'' in the following. One of the strongest manifestation of a functional relation between genes is Protein-Protein Interaction (PPI). Oti et al. \cite{Oti2006} demonstrated that exploiting PPIs can lead to novel candidate gene prediction. While many PPI network databases are available, such as \verb?Ingenuity Pathway Analysis? (Ingenuity® Systems, \url{www.ingenuity.com}) or \verb?STRING? \cite{Snel2000}, few studies have actually integrated such knowledge for biomarkers identification.\\
The PPI network was extracted using \verb?STRING?. The interactions include direct (physical) and indirect (functional) associations derived from various sources on the basis of both experimental evidence for PPIs as well as interactions predicted by comparative genomics and text mining. 

 \verb?STRING? uses a scoring system that is intended to reflect the evidence of predicted interactions. In the present study, we add the relevant functional partners with a score of at least 0.9 (which corresponds to a high-confidence network) to the set of differentially expressed genes.\\

In the following, our final set of genes will be called the "molecular signature" of the conditions to compare.\\

\subsection*{Step2: Pathway Analysis to define a robust \textit{a priori}}
Pathway Analysis aims at identifying the cellular processes involved in differentiating the two conditions to compare (cases and controls for instance) in order to enhance the interpretation of the molecular signature generated at step 1.\\

In practice, Pathway Analysis consists in determining whether the signature is enriched in pathway key actors. Given the $M$ genes measured on a microarray, our signature is defined as a subset of length $n$ of these genes, and a given pathway is defined as another subset of length $K$ of these genes. The probability of having $y$ genes in common between our signature and a pathway is given by the hypergeometric distribution: 

$$\mathbb{P}(Y=y)=\frac{\binom{n}{y}\binom{M-n}{K-y}}{\binom{M}{K}}.$$

This approach requires a pre-defined set of pathways to analyze. Several databases exist today and the Kyoto Encyclopedia of Genes and Genomes (KEGG \cite{Kanehisa2006}) is often taken as a reference.\\

The problem with the analysis of pathways is that they do not represent clearly distinct entities and two pathways can involve common genes and hence share a common biological information. Consequently a set of genes can be responsible for the positive results of several pathways.\\
In order to ease the interpretation of Pathway Analysis results, we propose to summarize the set of pathways found significant because of the same genes into a reduced set of "core pathways" (each core pathway represented by a set of pathways). In practice we apply a Hierarchical Clustering algorithm on a binary matrix, that contains the genes in row and the pathways in column, summarizing the membership of each gene to each pathway. Dissimilarity, which accounts for pairwise differences between the pathways, is assessed by using a binary metric, also known as the Jaccard distance. Finally, from this dissimilarity matrix we perform a Hierarchical Agglomerative Clustering using Ward's criterion.\\ 
Each core pathway defines a gene cluster which will be used in the next  step for  inferring the regulation network.

\subsection*{Step 3: Network Inference from biological prior knowledge}

The last step of the procedure consists in inferring a part of the regulation network using the transcriptomic data from all available conditions and from the  prior structure exhibited in  the pathway analysis. 
This last step relies on a Gaussian Graphical Model, taking advantage of the relation existing between conditional independence and the concentration matrix. Variables (i.e. genes)
$X^i$ and $X^j$ with  $i\neq j$ are  independent conditional on  all other
variables indexed by  $\mathcal{P} \backslash \{ i,j \}$,  if and only
if the  entry $({\boldsymbol\Sigma}^{-1})_{ij}$ is  zero.  The inverse
of        the       covariance       matrix        $\mathbf{K}       =
(K_{ij})_{(i,j)\in\mathcal{P}^2}=  {\boldsymbol\Sigma}^{-1}$, known as
the concentration matrix,  thus describes the conditional independence
structure of $X$.  Moreover, each  entry $K_{ij}$, is directly
linked      to      the      partial      correlation      coefficient
$r_{ij|\mathcal{P}\backslash\{i,j\}}$  between   variables  $X^i$  and
$X^j$.  The inferred graph is  described by  the  adjacency matrix defined by the nonzero elements of the concentration matrix. 

Merging different experimental conditions from transcriptomic  data is a common practice in GGM-based inference methods \cite{Toh2002}. This process enlarges the number of observations  available for inferring  interactions. However, GGMs
assume  that the  observed data  form an  independent  and identically distributed (i.i.d.) sample. But assuming that the merged  data is drawn from a single
Gaussian  component  is  obviously   wrong,  and  is  likely  to  have
detrimental side effects in the estimation process.

In  this  paper, we  assume that  the  distributions  of  these   different conditions  have  strong commonalities. We thus propose to  estimate these  graphs jointly. In this context, the multi-task learning provides a convenient framework \cite{Chiquet2010}. This approach allows to overcome  the difficulties arising from  the scarcity of data in each experimental condition by coupling the estimation problems.

Moreover we take advantage of the previously  exhibited gene clusters to put some prior structure on the inferred network.  We assume that the edges of the graph defined by  the nonzero elements of the concentration matrix 
are distributed among  a set  $\mathcal{Q}=\{1,\dots,Q \}$  of given overlapping 
clusters.   For any
gene  $i$, the  indicator variable $Z_{iq}$ is equal  to $1$ if $i\in
q$ and  $0$ otherwise, hence  describing to which cluster the  gene $i$
belongs.  The structure of the graph is thus described by the matrix 
   $\mathbf{Z}     = (Z_{iq})_{i\in\mathcal{P},q \in \mathcal{Q}}$.

Thus the group structure over  the $Q$ gene clusters  and the structure over the $2$ conditions lead  to estimating  the $2$ concentration matrices which are the solutions of the following  penalized log-likelihood maximization problem:
\begin{equation}
      \max_{\mathbf{K}^{(c)}, c\in {1,2}} \sum_{c=1}^2
      {\mathcal{L}}\left(\mathbf{K}^{(c)} ; \mathbf{X} \right) 
      -   \lambda    \sum_{\substack{i,j\in\mathcal{P}\\   i\neq   j}} \rho_{\mathbf{Z}_i \mathbf{Z}_j} 
      \left\{\left(\sum_{c=1}^2      \left[K_{ij}^{(c)}\right]_+^2
        \right)^{1/2}               +               \left(\sum_{c=1}^2
          \left[K_{ij}^{(c)}\right]_-^2 \right)^{1/2}\right\},
\end{equation}
    where  $\left[u\right]_+  =  \max(0,u)$  and  $\left[u\right]_-  =
    \min(0,u)$ and the coefficients of the penalty are defined as:
  \begin{equation}
    \label{eq:rho_func}
    \rho_{\mathbf{Z}_i \mathbf{Z}_j} = 
    \left\{\begin{array}{lr}
      \displaystyle \sum_{q,\ell\in\mathcal{Q}} Z_{iq}Z_{j\ell}
      \frac{1}{\lambda_{\text{in}}}, & \text{if } i\neq j, \text{ and } q=\ell,\\[5ex] 
       \displaystyle \sum_{q,\ell\in\mathcal{Q}} Z_{iq}Z_{j\ell}
      \frac{1}{\lambda_{\text{out}}}, & \text{if } i\neq j, \text{ and } q\neq \ell, \\[5ex]
       1, & \text{otherwise},
    \end{array}\right.
  \end{equation}
and \begin{equation*}
  \mathcal{L}(\mathbf{K}^{(c)};  \mathbf{X}  ) 
  = \frac{n}{2} \log  \det(\mathbf{K}^{(c)}) -  \frac{n}{2} \text{Tr}(\mathbf{S^{(c)}}
  \mathbf{K}) +C,
\end{equation*}
with $\mathbf{S^{(c)}}$  the empirical covariance matrix relative to  condition $c$ and  $C$  a  constant  term.

The first part of the criterion  $\sum_{c=1}^2 {\mathcal{L}}\left(\mathbf{K}^{(c)} ; \mathbf{X}\right)$ consists in the sum of the log-likelihoods of the two concentration matrices given the observations.  The second part of the criterion is a  penalty, which considers two types of edges: edges between two genes belonging to the same core pathway are penalized with a coefficient  $1/ \lambda_{\text{in}}$ and edges between two genes which are never present together in a pathway are penalized  with a coefficient  $1 / \lambda_{\text{out}}$.  

The basic  form of the penalty  has been proposed in \cite{Ambroise2009} where it  has been shown via simulations that  the  knowledge of an existing group structure was indeed improving the estimation of  the non-zeros entries of the concentration matrix. 
The current form of the penalty  \cite{Chiquet2010} encourages  co-regulation of the same sign across the two conditions.   Specifically,  co-regulation encompasses up-regulation and down-regulation and the type of regulation is not likely to be inverted across assays: in terms of partial correlations, sign swaps are very unlikely.

% Results and Discussion can be combined.

\section{Application}
We apply our approach to the study of the response to chemotherapy in breast cancers. The pathologic Complete Response (pCR), defined as the absence of disease in both the breast and lymph nodes, is used as an early surrogate marker of treatment efficacy. Most studies on response to chemotherapy have considered breast cancer as a single homogeneous entity. However, it is a complex disease that has a molecular and cellular heterogeneity that should not be overlooked in such studies. Perou et al. \cite{Perou2000} were the first to report evidence for breast cancer tumor subtypes based upon gene expression profiles. Since the various tumor subgroups were re-evaluated and breast cancers have been classified into five molecular classes: luminal A and B, basal, ErbB2+, and normal-like subgroups \cite{Sorlie2003, Sotiriou2003}. These subtypes differ markedly in prognosis and in the repertoire of therapeutic targets they express \cite{Nielsen2004}. Within these five groups, basal-like breast cancers, quickly became a subtype of interest because of its features. Indeed it is associated with poor prognosis and characterized by its aggressive behavior.\\

Many reports have demonstrated that basal-like tumors are chemo-sensitive. This particular subtype of breast cancer correlated with significantly higher pCR rates when compared to other subtypes \cite{Liedtke2008}. Nevertheless, basal tumors harboring homogeneous clinical and pathologic features may exhibit highly variable response to chemotherapy. It raises new questions to biologists such as how to explain the differences between patients who achieve a pCR from those that do not (not-pCR), and which are the molecular mechanisms involved in chemo-sensitivity or resistance ?\\

We demonstrate that our approach based on Pathway Analysis and Network Inference can provides effective tools to address such questions of biological and clinical interests. To do so, we investigate the dataset proposed by Hess et al. \cite{Hess2006} including 133 breast cancer samples for which the pCR status is known. Gene expression profiling has been made based on the Affymetrix U133A microarray. Following the methodology described in Sorlie et al. \cite{Sorlie2003}, we identify 29 basal tumors from the 133 samples, divided into 15 pCR and 14 not-pCR.\\

Step 1 - Comparison of the gene expression profiles between the pCR and not-pCR samples yields about 100 genes with statistically significant differences (at a $10^{-3}$ level). We add 30 key regulatory genes identified from the STRING database. These 130 genes represent the molecular signature of the pCR in our 29 basal tumors.\\

\begin{figure}[!ht]
\begin{center}
\includegraphics[trim = 0 70mm 0 40mm ,scale=0.4,clip]{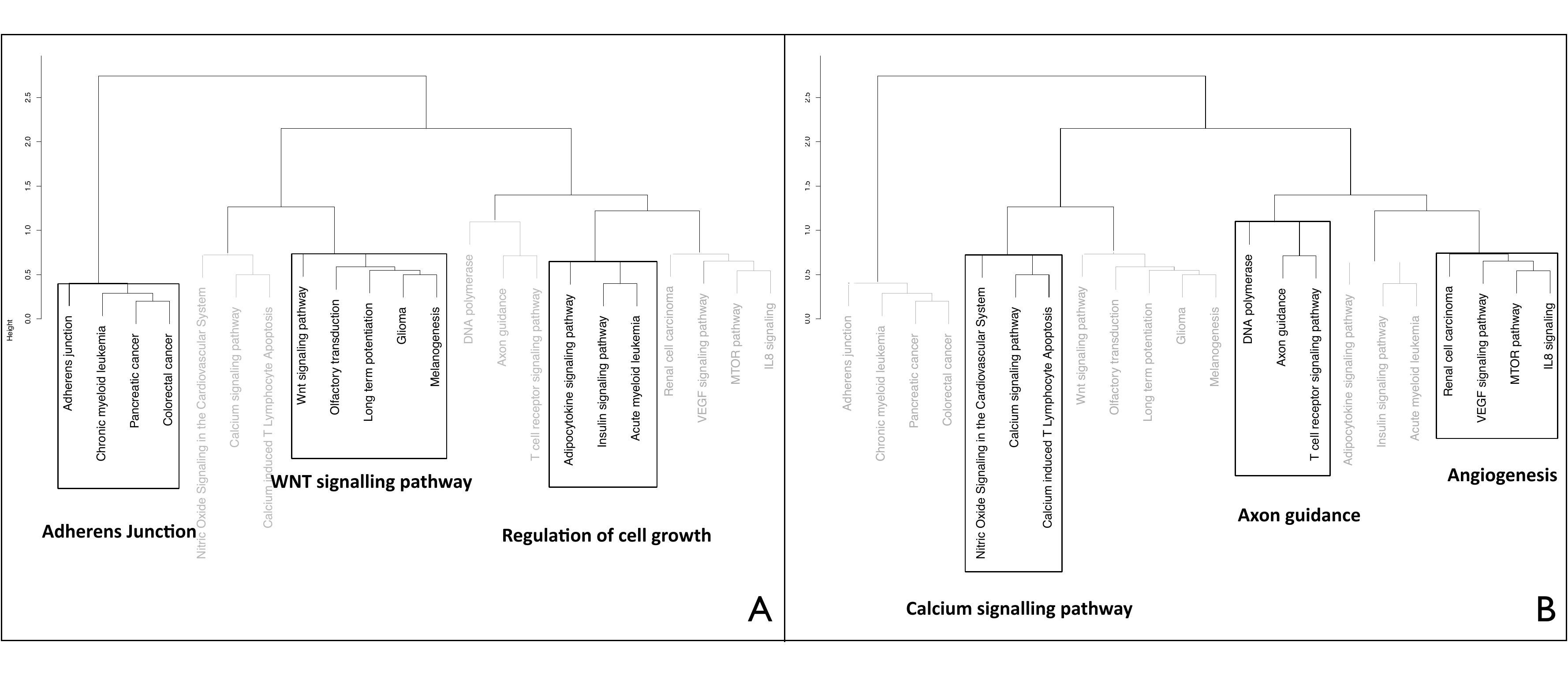}
\end{center}
\caption{\textbf{Core pathways.} From the Hierarchical Clustering we identify 6 core pathways. Figure \ref{PATHWAYS}-A highlights the core pathways related to tumor cell growth and proliferation mechanism and Figure \ref{PATHWAYS}-B highlights the core pathways associated with angiogenesis activity.}
\label{PATHWAYS}
\end{figure}

Step 2 -  The Pathway Analysis driven from the molecular signature identified 22 significant KEGG pathways (at a 5\% level). The Hierarchical Clustering of these pathways lead to 6 core pathways that summarize 6 main, distinct and clear biological processes involved in the mechanisms of response to treatment in basal tumors (Figure \ref{PATHWAYS}). For instance, one core pathway includes the VEGF, mTOR and IL8 signaling pathways, all associated with angiogenesis, the new blood vessel formation process required for tumor progression. Consequently, such results lead us to relate this core pathway to angiogenesis.\\

\begin{figure}[!ht]
\begin{flushleft}
\includegraphics[scale=0.55]{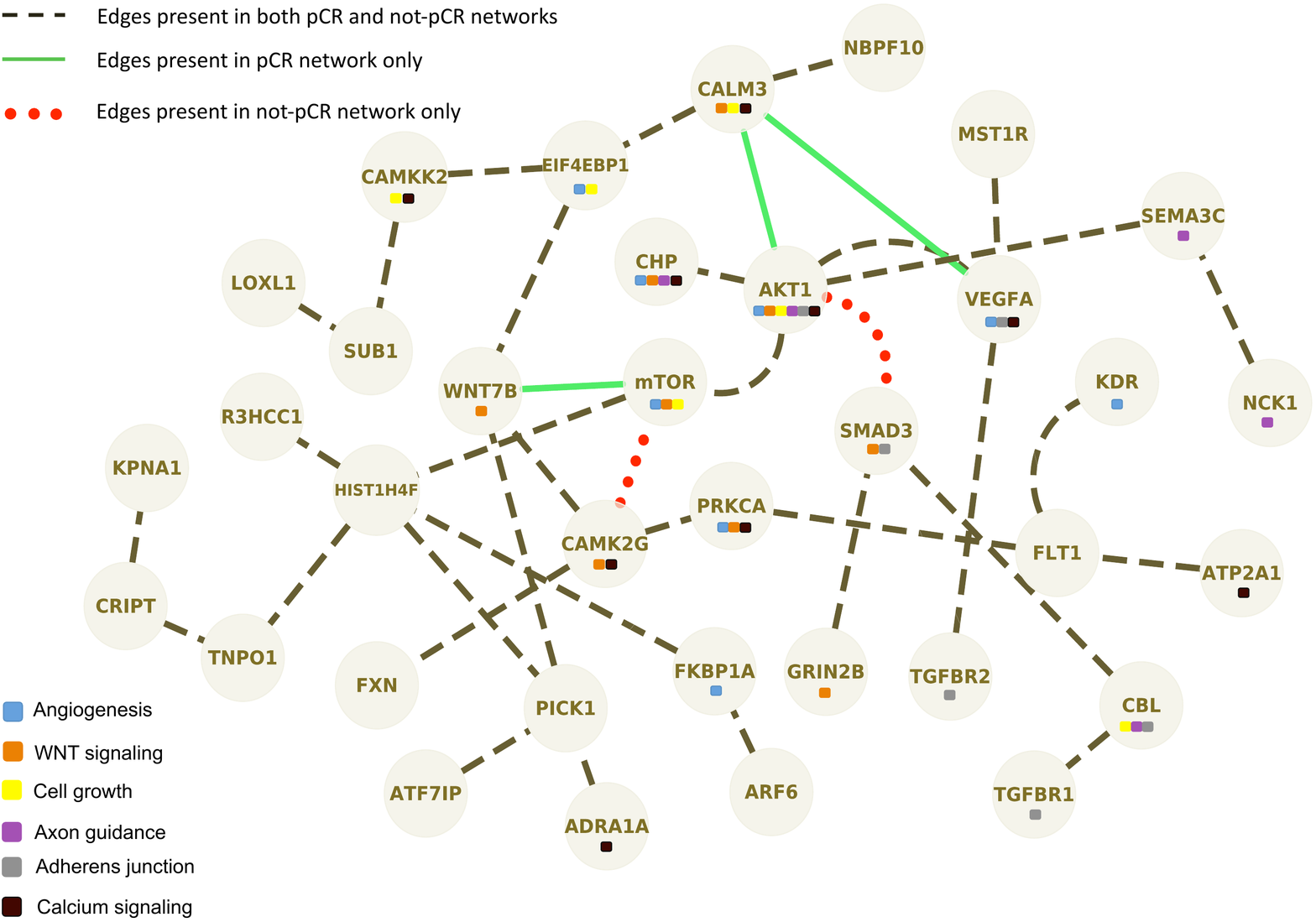}
\end{flushleft}
\caption{\textbf{Inferred graph from the Hess cohort.} The figure displays an enlarged view of a network subpart. The black dashed edges are present in both pCR and not-pCR networks. In red are the edges inferred only in the not-pCR network and conversely, the green edges are only inferred in the pCR network.}
\label{NETWORK}
\end{figure}

Step 3 - The 6 core pathways are taken as a biological prior for driving the Network Inference using the R package \verb?SIMoNe?. A particularly interesting network subpart is shown on Figure \ref{NETWORK}, wich includes many genes known to be associated with breast cancer (for instance: mTOR, WNT, VEGF, AKT1...).

In particular, it has been shown that the CALM3 gene regulates the activity of AKT1 in breast tumors \cite{Coticchia2009}. Our network suggests that this regulation occurs only in the pCR condition, and hence, may be broken for not-pCR tumors. 

\section{Discussion and conclusion}
 The ability to infer networks from high-throughput genomic data is one of the most promising applications of Systems Biology. In this paper, we propose a global framework to infer networks on the basis of a biological informative prior over network structures.  It has the advantage to reduce the space of possible network structures to investigate, and to propose a more relevant network in order to facilitate the interpretation of its edges. Network Inference is done using the SIMoNe algorithm developed by Chiquet et al. \cite{Chiquet2010} and based on a weighted Lasso criterion in a Gaussian Graphical Model framework.\\  

In this context, determining the biological prior to use is not a simple task. 
To this purpose, we define an informative prior using Pathway Analysis that allows, from molecular signature, to identify which are the main cellular mechanisms associated with a condition of interest. Many public domain repositories exist for storing biological pathways, each based on its own set of conventions. Thus pathways are not consistent among databases and the same pathway can present several definitions that vary greatly in quality and completeness \cite{Adriaens2008}. Pathway Analysis is hence database dependent. However our approach overcomes this limitation in two points: i) the definition of core pathways from the molecular signature makes our analysis less sensitive to the strict definition of pathways found in the databases and ii) it is possible to adjust the weighted penalization according to the confidence we have in each pathway definition.\\

The core of such an approach, combining both Pathway Analysis and Network Inference, is the selection of the most informative set of genes to discriminate between the two conditions (i.e. the molecular signature). The relevance of the biological findings at the end of the analysis greatly depends on this signature. In consequence, as a first step of our global Network Inference process, we propose an improved Differential Analysis. In particular, we aim to increase the reproductibility of the signature by removing potential outliers and identifying modules of genes via functional partners identification. It allows to include some well-studied genes (such AKT1) not detected in the differential analysis. However the noise inherent to PPI data is one of the main limitations of this approach. Indeed, only few protein interactions are confirmed by the HPRD (Human Protein Reference Database). These high noise levels reduce the accuracy of functional partners identification. By restricting our study to high-confidence score interactions, we focused on the most biologically relevant interactions. However it introduces a bias into our approach by selecting the most studied and documented proteins. Then, the functional partners identification step is highly dependent on the quality and the amount of data available.\\ 
The functional partners identification method still needs to be further improved and explored. In particular, we have to develop an evaluation strategy to better assess the contribution of such a method in identifying genes of interest.\\

Our application on breast cancer data highlights key regulations in cancer progression and response to treatment. The co-expression of AKT1 and CALM3 that occurs specifically in the group responding to chemotherapy is a good example. Such results can be a starting point of more important applications at the biological and clinical levels.\\

Finally, the regulation of gene expression is a complex process resulting from several regulatory mechanisms occurring at distinct steps of the biological system (miRNA, eQTL, genomic alterations and epigenetic factors for instance) and gene expression data enables us to understand a limited part of the whole system. In this context, the inference of networks from heterogeneous biological data, such as eQTL, is a very promising topic to develop in the near future.

%######################################## 
%\begin{acknowledgement}
\section*{Acknowledgement}
We thank Caroline Paccard, Matthieu Bouaziz, Fabrice Glibert, Julien Chiquet, Carene Rizzon, Claudine Devauchelle, Camille Charbonnier and Serguei Nabirotchkin for helpful discussions. We also thank Celeste Lebbe and Ilya Chumakov for their support.
%\end{acknowledgement}

\section*{Author Contributions}
Conceived and implemented the global process: MJ. Wrote the paper: MJ, MG and CA. Co-supervised the study: MG and CA.
\bibliography{BBC}

\end{document}